Inverse Thinking in Economic Theory: A Radical Approach to Economic Thinking

Jaime Gomez-Ramirez
Universidad Politécnica de Madrid
jd.gomez@upm.es

Abstract

The seriousness of the current crisis urgently demands new economic thinking that breaks the *austerity vs. deficit* spending circle in economic policy. The core tenet of the paper is that the most important problems that natural and social science are facing today are inverse problems, and that a new approach that goes beyond optimization is necessary. The approach presented here is radical in the sense that identifies the roots in key assumptions in economic theory such as optimal behavior and stability to provide an inverse thinking perspective to economic modeling of use in economic and financial stability policy. The inverse problem provides a truly multidisciplinary platform where related problems from different disciplines can be studied under a common approach with comparable results.

1. Introduction

The relationship between economic aggregates such as the unemployment rate or inflation, based on direct measurement of quantities is an inverse problem which is not well posed. The inverse problem is to infer the value of parameters of interest based on the direct measurement of observables. This form of inference is ill-posed in the sense that solutions to the problem may not exist, be multiple and be instable, that is, small errors in the measurements lead to large differences in the solution.
In order to be able to build on this fundamental but poorly understood insight we need to critically explore pillars of orthodox economic theory like the rational expectation hypothesis, market efficiency and stable equilibrium, providing new avenues and hands-on alternatives for a more realistic and resilient economic models.
More specifically, we claim that the new generation of economists must be aware of the necessity to deal with the next three points:

1. Develop a new theoretical framework based on the inverse problem theory that will work as a scaffold where key topics like stability, causality or predictability in economic models.
2. Establish a multidisciplinary program to gather and analyze data relative to cognitive and ecological features of interest for economic system modeling.
3. Develop new modeling techniques, mathematical and computational, including complexity theory tools like networks to model issues related robustness.

2. Four problems in classical economic modeling

Economies are complex man-made systems where organisms and markets interact according to motivations and principles not entirely understood yet. The increasing dissatisfaction with the postulates of traditional economics i.e. perfectly rational agents, interacting through efficient and stable markets, has created new incentives for different approaches in economics. For example, behavioral economics (Akerlof:2002) builds on cognitive and emotional models of agents, neuroeconomics (Glimcher:2003) addresses the neurobiological basis of valuation of choices, and evolutionary economics (Arthur:1999) that studies economies as a complex evolutionary system, composed of agents that adapt to endogenous patterns out of equilibrium regions. Social science, and in particular economics, is undergoing a decisive historical moment. A new approach in economic theory able to palliate the dissatisfaction with core tenets in classical economics is sorely needed. In what follows we briefly describe four problems –agent-economy separation, causality, equilibrium and perfectly rational maximizers- that need to be specifically addressed.

2.1. Agent-Economy separation

Orthodox economic thinking draws a neat line between agents and the economy they perform, to that effect microeconomics and macroeconomics are publicized as the suitable disciplines to address the issues related to the former and the last respectively. Samuelson changed for ever the study of economics, which became strongly mathematical by borrowing methods and insights from statistical physics. University departments of economics are an institutionalized embodiment of his vision. However, the idiosyncrasy of economic systems composed of participants endowed with cognitive and emotional characteristics clearly calls into question the analogy between physics and economics. Economic agents are not merely reactive but have an anticipatory nature. Realistic economic systems try to predict the outcomes of their actions and of those of other agents, and in doing so they modify the structure itself that they are trying to cope with or outperform, as in financial markets. The forward-looking capacity of biological and social systems is an additional layer of complexity that statistical mechanics applications lack.

2.2. Causality

Econometric models have mainly focused on models of exchange and allocation, neglecting critical aspects in economies' dynamics like production or credit money. This apparent lax approach has its root in modeling economic systems "as if" (Friedman:1966) having an equivalent ontological status with the rest of material systems that form the "book of nature that is written in the language of mathematics", as Galileo poetically stated. However, the "unreasonable effectiveness of mathematics in physics" relies upon the existence of invariance principles, observed in quantities that remain constant regardless the system's motion (Wigner:1960). Economies are not conservative systems, assets are created *ex novo*, as in the process of money creation through credit; and disappear, enacted by the innovative entry of new technologies. This is the disruptive force that sustains economic growth or Schumpeter's "creative destruction" (Schumpeter:1994). Thus, variance principles and general laws in economics are more a desideratum than a factum. The Lucas critique (Lucas:1976) is still valid, even though the approaches that followed like DGSE forgo realism for mathematical virtuosity, macroeconometric models are not invariant to economic policy, or simply put, a policy experiment may modify the structural parameters that is trying to predict. The causal thinking in economics, which mainly relies on "the genetic-causal tradition" (Cowan:1996) must be explored with fresh insight.

2.3. Equilibrium

In Newtonian physics and classical thermodynamics the laws that govern the behavior of single particles suffice for the study of assemblies of particles, but this "methodological bonus" comes with a cost, they are theories of equilibrium, and therefore not apply to far-from-equilibrium systems. Equilibrium is a terminal state that is reached when all process have stopped. Biological and social systems are open systems that transition between different regimes, so non stationarity and volatility have to be taken into account. The participants in economic systems are heterogeneous. They are equipped with different informational processing capabilities, different systems of belief and ways to internalize the world. Thus, their interactions may produce novel dynamics and patterns whose understanding lie outside the basic assumptions necessary to establish an equilibrium, such as perfect competition or perfect knowledge of all the participants. Participants in an economy cannot be merely averaged out as particles in a gas, as they try to adapt to a complex, highly coupled and self-referential eco-systems. One unique stable equilibrium cannot be taken for granted, instead multiple equlibria and qualitative regime shifts are more realistic signatures of economic systems. Mathematical tractability which is undoubtedly a good in itself, should not lead us to overlook this fact. In order to deal with instability-prone phenomena like increasing returns or leverage we may need to design new regulatory mechanisms other than homeostasis-based.

2.4. Perfectly Rational Maximizers

Selfishness is one of the most reliable characteristics of human beings, but that empirically testable statement does not preclude other forms of economic interaction. The egalitarian assumption that all the participants in an economic system are utility maximizers has a practical motivation, it likens economic modeling to an optimal problem. In doing so, no attention is paid to stability because optimal systems are in general stable (Astrom:2006).
From a control theory perspective the Keynesian and the Friedman-Lucas approaches share the idea of the omniscient controller. In a Keynesian's stance, the state is the controller that aims to stabilize the system, and not the mere input as in, for example, deficit spending. On the other hand, the Friedman-Lucas approach relegates the state to a trivial role, that to provide a constant input e.g. money supply, in order to have n rather controllers than 1.
However, in this decentralized approach, there is implicitly a controller –the economist- who defines the objective function: utility maximization, which is typically the wealth, or even better the logarithm of the wealth. Orthodox economic policies have a Panglossian view of economic participants with identical interests competing in the best possible world, which only exists in textbooks of macroeconomic modeling. We need to take a more pessimistic and honest view: decision makers behavior may largely deviate from optimal criteria, utility functions may not be computable, and people may make systematic errors in predicting the future driving the system to multiple equilibria and instability.

3. A new outlook to economic modeling

The core tenet of this work is that the most important problems that natural and social science are facing today are inverse problems, and that a new approach that goes beyond optimization, that takes into account the subjective knowledge of the agent is necessary. The inverse problem is ubiquitous in science; molecular structural reconstruction; metabolic network construction; biomedical imaging reconstruction of tumors; reconstruction of the internal structure of the earth; cognitive modeling or option-pricing model in financial markets, are all examples of the inverse problem (Groetsch:1993).

3.1. Inverse Thinking

To explain, for example, how an increase in interest rate will cause aggregate demand to decline, we have to study causality. However we can't trace linear causal chains in complex economic systems such as national economies. Other forms of causality like circular, downward or multiple causality need to be taken into account. We need to foster an "inverse thinking" approach to study causality and related issues e.g. predictability. Predictability can be seen as a forward model that takes a cause $C$ into an effect $E$. The forward problem or modelization allows to make predictions on the value of observables, and the inverse problem uses those predictions to infer the values of the parameters that characterize the system. Thus the inverse problem entails the existence of a forward problem to be inverted.
Causality as an inverse problem and needs to be studied as such. Thus, very schematically *inv(fwd(C))=C'* states that if $C=C'$ the inverse problem has an unique solution so we have a linear causality, otherwise, $C \neq C'$, the inverse problem is ill-founded because two causes $C$ and $C'$ produce the same effect $E$. Within an inverse thinking approach we don't try to solve an inverse problem that is ill-founded, as for example in *C*(interest rate increase) causes *E*(aggregate demand decrease), but we collect all the forward models, that is, all the predictors that are not falsified by data. Thus we build a set of possible causes or forward model solutions rather than one prediction. The rationale behind this liberal approach to causality is that in order to deal with uncertainty and volatility we need to have a multiple and adaptable hypothesis scheme.

Inverse thinking challenges the orthodox linear causal chain of reasoning, dramatic system disruptions do not necessarily need large perturbation to occur. The approach is speculative and Popperian in the sense that proposes axioms and predictions that can be refuted. The inverse thinking approach will help to discover the design principles for robust systems in the face of uncertainty and complexity.

3.2 Bio-inspired homeostasis

Conceptual clarification and epistemic cleansing in the misuse of analogies between physics and economy drawn is a requisite for developing comprehensible economic knowledge, independently of the particular domain in which it is used. Economies are not mere exchange systems, the transformation of labor and energy into tangible goods or production is also part of real economic systems. From a thermodynamic point of view, the term economic equilibrium is an oxymoron. Equilibrium is a frozen state of no change, where no macroscopic flows exist. Thus, it is a concept more likely to apply to closed or isolated systems rather than to real economies.

The concept of non equilibrium steady state needs to be carefully addressed as a process that needs to be maintained by the interplay of non zero flows of energy or matter, rather than a final state as equilibrium. Future research should try to develop scenarios with *n* different equilibria with different dynamic regimes. The inverse thinking approach may provide the necessary mindset to strengthen our understanding of multiple equilibria, as it natively deals with a scenario of multiple and instable solutions, including "bubble solutions" or large deviations from equilibrium. To turn these theoretical insights into real applications, we need to develop biologically inspired models of stability mechanisms, in particular homeostasis (Simon:2002). A new view of homeostasis that deals with integrated outcomes rather than single signals is already being promoted in biomedical applications (Hotamisligil:2006).

3.3. Utility

Non parametric estimators can be arbitrarily well approximated, this property is called consistency and it is the major reason for the popularity of model-free estimators. Consistency guarantees that, for a sufficiently big training data set, it is possible to achieve an optimal performance for any inference learning task. Thus, there is apparently a causal connection between consistency and optimality. However, non parametric estimators are optimal because they are consistent, but consistency is an asymptotic property. In real problems, the training data set cannot be assumed to be arbitrarily big, data samples may be small or have dispersed distribution. This point has direct consequences in economic policy. The standard approach in optimal economic policy, which is mainly concerned with performing a gradient ascent on utility or wealth maximization, should be critically revisited. Modeling all economic systems as optimal utility maximizers is at least questionable using empirical standards, as it depends on consistency which is an asymptotic property. The Allais' paradox (Quiggin:1993) can be seen as a reflect of the empirical limitation of using optimal methods when consistency may not necessarily apply. Indeed the same person that prefers an apple to an orange, may prefer one dollar and one orange to one dollar and one apple. The idea of using an optimization algorithm to forecast people's choice as represented in the utility function is untenable.

We need to transcend the idea of one consistent and optimal criterion for all the participants behavior, proposing a multifold space of subjective utility models, represented by a *n* dimensional set of subjective expected utility (SEU). In order to take into account how individual SEU maximization affects and is affected by the aggregate SEU, each SEU is considered within an ecological perspective. It ought to be remarked that SEU is a non-convex utility function, so it is not the same as Bayesian utility maximization.

The inverse problem must start by introducing all the available information or bias to constrain the model space of predictors represented by subjective utility functions. Bias or a priori knowledge, works as a selection mechanism over the space of falsifiable models, which in turn are tested against the data, that is, those that do not predict data within an established criterion are discarded or falsified in Popperian parlance. Thus, it is within the inverse thinking perspective that we can start understanding the computational goal(s) the economic systems under study.

4. Discussion

Our zeitgeist is that of deregulation and credit. The political links of special interest groups were not the only driving force that pushed economies in this state of affairs. At a more fundamental level, deregulation can be seen as a natural outcome of the increase in complexity in the world network economy that globalization has bring about. In the 70's stabilization policies designed to for example control in inflation and unemployment using optimal control theory, did not succeed (Kalchbrenner:1977), (Prescott:1977). National economies became too complex to be regulated with that toolset. Monetarist economists' credo that, the system itself could self-regulate without any need to an external controller, that could even worsen the economy, was championed by policy makers. The view that complexity was beneficial for its stability was widely assumed in ecology studies, until the 70s when the mathematical biologist Robert May demonstrated that complexity may engender instability (May:1972). Apparently, economists were unaware of this work, as the standard economic modeling of optimal forecasters in stable equilibrium remained relatively undisputed until the 2008 financial crisis.

By focusing in the inverse problem we may reduce the isolation of economic theories from the natural science, as it provides an ideal platform where related problems from different disciplines can be studied under a common approach with comparable results. New means to study quantities that are not directly observable like utility, drawing causal maps with measurements gathered from financial economics, need to be scrutinized. By adopting an inverse perspective, we contribute to the creation of new ways of economic thinking and to the discovery of new tools to tackle inverse problems. In doing so, the economic discipline will be emancipated from its ties to XIX century physics to enter into a new paradigm shift. A new picture of stability will emerge from the study of the cognitive and ecological dimensions in economic systems, grounded in the conceptual and methodological apparatus provided by the inverse problem theory.